\documentclass[nofootinbib,preprintnumbers,superscriptaddress,notitlepage]{revtex4}

\usepackage{epsfig}
\def\beq{\begin{equation}}
\def\eeq{\end{equation}}
\def\bea{\begin{eqnarray}}
\def\eea{\end{eqnarray}}

\begin{document}
\title{Improved vacuum stability in a 5D extra dimension model
}

\author{Lu-Xin~Liu}
\email[Email: ]{luxin.liu9@gmail.com}
\affiliation{National Institute for Theoretical Physics; School of Physics, University of the Witwatersrand, Wits 2050, South Africa}
\author{A.~S.~Cornell}
\email[Email: ]{alan.cornell@wits.ac.za}
\affiliation{National Institute for Theoretical Physics; School of Physics, University of the Witwatersrand, Wits 2050, South Africa}

\begin{abstract}
In this paper we study the renormalization effects of the quark flavor mixings and the Higgs self-coupling in a five dimensional model where the boson fields are propagating in the bulk whilst the matter fields are localized to the brane. We first explore the evolution behaviors for the Cabibbo-Kobayashi-Maskawa matrix in this scenario. Then, in light of the recent LHC bounds on the Higgs mass, we find that the Higgs self-coupling evolution has an improved vacuum stability condition, which is in contrast with that of Standard Model and the Universal Extra Dimension scenario, where the theory has a much lower ultraviolet cut-off.
\end{abstract}

\date{10 July, 2012}
\preprint{WITS-CTP-91}
\maketitle


\section{Introduction}\label{sec:1}

\par With the Large Hadron Collider (LHC) now up and running, physicists have begun to explore the realm of new physics that may operate at the TeV scale, where such physics will lead to greater understanding of the mechanism that breaks electroweak (EW) symmetry and generates the masses of all known elementary particles. Consider the quark sector of the SM. We have ten experimentally measurable parameters, i.e. six quark masses, three mixing angles, and one phase. A completely satisfactory theory of fermion masses and the related problem of mixing angles is certainly lacking at present. Recall that in order to explore the physics at a high energy scale we use renormalisation group equations (RGEs) as a probe to study the momentum dependence of the Yukawa couplings, the gauge couplings, and the Cabibbo-Kobayashi-Maskawa (CKM) matrix elements themselves. As such we can consider one of the primary goals of the LHC as being to uncover any new dynamics within the TeV range. Instead of assuming the RGE goes from the $M_Z$ scale up to the GUT scale ($10^{15}$ GeV) by using the $SU_C(3)\times SU_L(2) \times U_Y(1)$ symmetry, we know that models with extra dimensions may bring down the unification to a much lower energy scale.

\par On the other hand, recall that in the Standard Model (SM) of elementary particles and their interactions, the Higgs field is integral to this mechanism and is the last missing and the most important piece. Recently, the ATLAS and CMS collaborations presented new bounds on the mass of the SM Higgs boson within a range 115.5GeV -127GeV at 95\% confidence level \cite{LHCdata}. Direct observational constraints on the Higgs boson are significant for its detection as well as its final discovery. While the search for the Higgs boson is one of the primary goals of the LHC, the experimental collaborations are also intensively searching for signs of new physics beyond the SM. Among the most popular models describing new physics within the reach of the LHC are extra dimensional models, where the SM particles can either all have access to the extra dimension bulk space or be confined on the 3+1 dimensional brane \cite{Cornell:2011fw, Deandrea:2006mh, Blennow:2011mp, Muck:2001yv, Dienes:1998vg}.

\par The RGEs are an important tool for the search of the properties of the quark masses, the CKM matrix, and the Higgs self coupling at different energy scales. It is therefore of great interest to have an implementation of extra dimensional models in studying these RGEs. In this paper we consider the Universal Extra Dimension (UED) model with a single compactified extra dimension, with the SM fermions confined on the 3+1 dimensional brane, whilst the gauge fields and Higgs can propagate in the bulk. The paper is organized as follows. In Section \ref{sec:2}, based on the brane localized matter fields scenario, we first develop the one-loop RGEs for the Yukawa couplings, the CKM matrix, and the Higgs quartic coupling. In section 3 we shall quantitatively analyze the evolution of these RGEs from low energies up to the unification scale. The scale dependence of the mixing angles as well as the CP violation measure $J$ will be plotted. In addition, motivated by the recent hints of the SM Higgs at the LHC, it is very natural for us to investigate the vacuum stability bounds of the Higgs boson within this alternative framework. As such, in section \ref{sec:4}, within the ATLAS and CMS Higgs bounds, we quantitatively analyze the Higgs self coupling evolution from the EW scale up to the unification scale and exploit its evolution behaviors for different compactification radii. The scalar potential is found to be positive definite and the EW vacuum can be stabilized at high energies. The ultraviolet cut-off of the theory may then be pushed up all the way to the gauge unification scale. The last section is devoted to a summary and our conclusions.


\section{The Evolution Equations}\label{sec:2}

\par The RGEs of the five dimensional UED model with one single extra dimension, compactified on a circle of radius $R$ with a $Z_2$ orbifolding, has been extensively discussed in Refs. \cite{Cornell:2011ge,Cornell:2010sz,Liu:2011gr}. We shall now consider the case of brane localized matter fields in a five dimensional model. Following the notation in Ref. \cite{Cornell:2010sz}, the kinetic terms of the matter fields and the scalar doublet have the following forms in this five dimensional model:
\bea
{\cal L}_{Quarks} &=& i{{\bar q}_L}(x){\gamma ^\mu }{D_\mu }{q_L}(x) + i{{\bar u}_R}(x){\gamma ^\mu }{D_\mu }{u_R}(x) + i{{\bar d}_R}(x){\gamma ^\mu }{D_\mu }{d_R}(x)\nonumber \; , \\
{\cal L}_{Leptons} &=& i{{\bar L}_L}(x){\gamma ^\mu }{D_\mu }{L_L}(x) + i{{\bar e}_R}(x){\gamma ^\mu }{D_\mu }{e_R}(x)\nonumber \; , \\
{\cal L}_{Higgs} &=& \int\limits_0^{\pi R} {dy} {({D_M}\Phi (x,y))^\dag }{D^M}\Phi (x,y) \; , \label{eqn:1}
\eea
where the extra dimension of the matter fields action has been integrated out using the delta function $\delta (y)$. This is done explicitly where the covariant derivative terms have the form
\bea
{D_\mu }{q_L}(x) &=& \left\{ {\partial _\mu } + ig_3^{5}{G_\mu }(x,0) + ig_2^{5}{W_\mu }(x,0) + i\frac{{{Y_Q}}}{2}g_1^{5}{B_\mu }(x,0)\right\}{q_L}(x)\nonumber \; , \\
{D_\mu }{u_R}(x) &=& \left\{{\partial _\mu } + ig_3^{5}{G_\mu }(x,0) + i\frac{{{Y_u}}}{2}g_1^{5}{B_\mu }(x,0)\right\}{u_R}(x)\nonumber \; , \\
{D_\mu }{d_R}(x) &=& \left\{{\partial _\mu } + ig_3^{5}{G_\mu }(x,0) + i\frac{{{Y_d}}}{2}g_1^{5}{B_\mu }(x,0)\right\}{d_R}(x)\nonumber \; , \\
{D_\mu }{L_L}(x) &=& \left\{{\partial _\mu } + ig_2^{5}{W_\mu }(x,0) + i\frac{{{Y_L}}}{2}g_1^{5}{B_\mu }(x,0)\right\}{L_L}(x)\nonumber \; , \\
{D_\mu }{e_R}(x) &=& \left\{{\partial _\mu } + i\frac{{{Y_e}}}{2}g_1^{5}{B_\mu }(x,0)\right\}{e_R}(x) \; . \label{eqn:2}
\eea
However, for the scalar field, which is in the bulk, its covariant derivative is
\beq
{D_M}\Phi (x,y) = \left\{{\partial _M} + ig_2^5{T^a}W_M + \frac{i}{2}g_1^5{B_M}\right\}\Phi (x,y)\; , \label{eqn:3}
\eeq
where the gauge fields $G_M (x,y) \; \displaystyle \left( G_M^A \frac{\lambda^A}{2} \right)$, $W_M (x,y) \; \displaystyle \left( W_M^a \frac{\tau^a}{2} \right)$ and $B_M (x,y)$ refer to the $SU(3)$, $SU(2)$ and $U(1)$ gauge groups respectively. Note also the five dimensional gauge coupling constants $g_3^5$, $g_2^5$ and $g_1^5$ are related to the four dimensional SM coupling constants (up to a normalization factor) by $\displaystyle g_i = \frac{g_i^5}{\sqrt{\pi R}}$, and similarly for the quartic coupling $\displaystyle \lambda = \frac{\lambda^5}{\sqrt{\pi R}}$. From a four dimensional view point, every field in the bulk will then have an infinite tower of Kaluza-Klein (KK) modes, with the zero modes being identified as the SM state. The five dimensional gauge fields have the generic form $A_M = (A_\mu, A_5)$ and the fifth component of the gauge bosons $A_5(x,y)$ is a real scalar and does not have any zero mode, transforming in the adjoint representation of the gauge group. As such, the five dimensional KK expansion of the gauge fields and the scalar field become
\bea
{A_\mu }(x,y) &=& \frac{1}{{\sqrt {\pi R} }}\left\{ A_\mu ^0(x) + \sqrt 2 \sum\limits_{n = 1}^ \propto  {A_\mu ^n(x)\cos \left(\frac{{ny}}{R}\right)} \right\}\nonumber \; , \\
{A_5}(x,y) &=& \sqrt {\frac{2}{{\pi R}}} \sum\limits_{n = 1}^ \propto  {A_5^n(x)\sin \left(\frac{{ny}}{R}\right)} \nonumber \; , \\
\Phi (x,y) &=& \frac{1}{{\sqrt {\pi R} }}\left\{ \Phi (x) + \sqrt 2 \sum\limits_{n = 1}^ \propto  {{\Phi _n}(x)\cos \left(\frac{{ny}}{R}\right)} \right\} \; . \label{eqn:4}
\eea

\par After integrating out the compactified dimension, the four dimensional effective Lagrangian, Eq. (\ref{eqn:1}), contains interactions involving the zero mode and the KK modes. As expected, the values of physical parameters, such as Yukawa couplings and gauge couplings, do not run in the old fashion. For the one-loop diagrams of the Yukawa couplings,
at each KK excited level, the contributions from the KK tower corresponding to the fields in Eqs.(\ref{eqn:3},\ref{eqn:4}) exactly mirror those of the SM field ground states, plus new contributions from interactions due to the fifth component of the vector fields as shown in Ref. \cite{Cornell:2010sz}. If we consider our five dimensional theory as effective up to a scale $\Lambda$, then between the scale $R^{-1}$ (the compactification scale of the single flat extra-dimension), where the first KK states are excited, and the cut-off scale $\Lambda$, there are finite quantum corrections to the Yukawa and gauge couplings from the $\Lambda R$ number of KK states. As a result, once the KK states are excited, these couplings exhibit power law dependencies on $\Lambda$. This can be illustrated if $\Lambda R \gg 1$ (to a very good accuracy), the generic four dimensional beta function has been shown to have a power law evolution. Explicitly, the beta functions of the Yukawa couplings are found as follows (note that there are no contributions from the KK excited states of the fermions to the Yukawa couplings in this five dimensional model)
\bea
16{\pi ^2}\frac{{d{Y_U}}}{{dt}} &=& \beta _U^{SM} + \beta _U^{5D}\nonumber \; , \\
16{\pi ^2}\frac{{d{Y_D}}}{{dt}} &=& \beta _D^{SM} + \beta _D^{5D} \; , \label{eqn:5}
\eea
with
\bea
\beta _U^{5D} &=& {Y_U} \cdot 2(S(t) - 1)\left[ - \left(8g_3^2 + \frac{9}{4}g_2^2 + \frac{{17}}{{20}}g_1^2\right) + \frac{3}{2}(Y_U^\dag {Y_U} - Y_D^\dag {Y_D})\right]\nonumber \; , \\
\beta _D^{5D} &=& {Y_D}2(S(t) - 1)\left[ - \left(8g_3^2 + \frac{9}{4}g_2^2 + \frac{1}{4}g_1^2\right) + \frac{3}{2}(Y_D^\dag {Y_D} - Y_U^\dag {Y_U})\right] \; . \label{eqn:6}
\eea
In the lepton sector, we have
\beq
16{\pi ^2}\frac{{d{Y_E}}}{{dt}} = \beta _E^{SM} + \beta _E^{5D}\; , \label{eqn:7}
\eeq
and
\beq
\beta _E^{5D} = {Y_E} \cdot 2(S(t) - 1)\left[ - \left(\frac{9}{4}g_2^2 + \frac{9}{4}g_1^2\right) + \frac{3}{2}Y_E^\dag {Y_E}\right]\; , \label{eqn:8}
\eeq
where $Y_E = diag(y_e, y_\mu, y_\tau)$, and the SM beta functions $\beta_U^{SM}$, $\beta_D^{SM}$, and $\beta_E^{SM}$ can be found in the Appendix. Note that here $t = \ln \left( \mu / M_Z \right)$ is the energy scale parameter, where we have chosen the $Z$ boson mass as the renormalization point, and $S(t) = e^t M_Z R$. That is, when the energy $\mu  = 1/R$, or $S(t)=1$, the whole beta function reduces to the normal beta functions (as when $\mu  \le 1/R$ there are no KK modes, so it is reasonable that the physics be described by the normal SM beta functions).

\par By imposing the unitary transformation on both sides of the evolution equations of $Y_U^\dag Y_U$ and $Y_D^\dag Y_D$, we have the RGEs for the eigenvalues of the square of these Yukawa coupling matrices as follows
\bea
16{\pi ^2}\frac{{d{f_i}^2}}{{dt}} &=& {f_i}^2\left[2T - 2{G_U} + 3(2S(t) - 1){f_i}^2 - 3(2S(t) - 1)\sum\limits_j {{h_j}^2} {\left| {{V_{ij}}} \right|^2}\right]\nonumber \; , \\
16{\pi ^2}\frac{{d{h_j}^2}}{{dt}} &=& {h_j}^2\left[2T - 2{G_D} + 3(2S(t) - 1){h_j}^2 - 3(2S(t) - 1)\sum\limits_i {{f_i}^2} {\left| {{V_{ij}}} \right|^2}\right] \; , \label{eqn:88}
\eea
along with
\bea
16{\pi ^2}\frac{{d{y_a}^2}}{{dt}} &=& {y_a}^2\left[2T - 2{G_E} + 3(2S(t) - 1){y_a}^2\right]\; , \label{eqn:9}
\eea
where $T = Tr\left[ 3 Y_U^\dag Y_U + 3 Y_D^\dag Y_D + Y_E^\dag Y_E \right]$, $G_U = \displaystyle \left( 2 S(t) - 1 \right) \left( 8 g_3^2 + \frac{9}{4} g_2^2 + \frac{17}{20} g_1^2 \right)$,\\ $G_D = \displaystyle \left( 2 S(t) - 1 \right) \left( 8 g_3^2 + \frac{9}{4} g_2^2 + \frac{1}{4} g_1^2 \right)$, and $G_E = \displaystyle \left( 2 S(t) - 1 \right) \left(  \frac{9}{4} g_2^2 + \frac{9}{4} g_1^2 \right)$.

\par Consequently, the CKM running of the quark flavor mixing matrix in the charged current is governed by
\bea
16{\pi ^2}\frac{{d{{\left| {{V_{ij}}} \right|}^2}}}{{dt}} &=& (2S(t) - 1)\left\{ 3{\left| {{V_{ij}}} \right|^2}\left({f_i}^2 + {h_j}^2 - \sum\limits_k {{f_k}^2} {\left| {{V_{kj}}} \right|^2} - \sum\limits_k {{h_k}^2} {\left| {{V_{ik}}} \right|^2}\right)\nonumber \right. \\
&& \hspace{1cm} - 3{f_i}^2\sum\limits_{k \ne i} {\frac{1}{{{f_i}^2 - {f_k}^2}}} \left(2{h_j}^2{\left| {{V_{kj}}} \right|^2}{\left| {{V_{ij}}} \right|^2} + \sum\limits_{l \ne j}{h_l}^2{V_{iklj}}\right) \nonumber \\
&& \hspace{1cm} \left. - 3{h_j}^2\sum\limits_{l \ne j} {\frac{1}{{{h_j}^2 - {h_l}^2}}} \left(2{f_i}^2{\left| {{V_{il}}} \right|^2}{\left| {{V_{ij}}} \right|^2} +\sum\limits_{k \ne i}{f_k}^2{V_{iklj}}\right)\right\}\; , \label{eqn:10}
\eea
where $V_{ijkl} = 1 - \left| V_{il} \right|^2 - \left| V_{kl} \right|^2 - \left| V_{kj} \right|^2 - \left| V_{ij} \right|^2 + \left| V_{il} \right|^2 \left| V_{kj} \right|^2 + \left| V_{kl} \right|^2 \left| V_{ij} \right|^2$.

\par The Higgs quartic coupling $\lambda$, which gives the mass of the Higgs scalar, is given as $\displaystyle \frac{\lambda}{2} \left( \phi^\dag \phi \right)^2$. A calculation of the KK mode contributions to its one-loop beta function in our model is completely analogous to that of the five dimensional UED model in Ref. \cite{Cornell:2011ge} (however, there are no contributions from the KK excited states of the fermion fields to the Higgs self coupling four points functions in the brane localized matter fields model). As a result, the evolution of the Higgs quartic coupling is given by the beta function as follows
\beq
16{\pi ^2}\frac{{d\lambda }}{{dt}} = \beta _\lambda ^{SM} + \beta _\lambda ^{5D}\; , \label{eqn:11}
\eeq
where
\beq
\beta _\lambda ^{5D} = (S(t) - 1)\left\{ {12{\lambda ^2} - 3\left( {\frac{3}{5}g_1^2 + 3g_2^2} \right)\lambda  + \left( {\frac{9}{{25}}g_1^4 + \frac{6}{5}g_1^2g_2^2 + 3g_2^4} \right)} \right\}\; . \label{eqn:12}
\eeq
Note that when comparing the Yukawa coupling beta function Eq. (\ref{eqn:6}) with these in Ref. \cite{Bhattacharyya:2002nc}, there are no KK modes associated with the fermion trace term, which actually originates from the fermion loop renormalization of the Higgs field and thus is absent in this brane localized fermion field scenario. Also, the pure gauge terms in Eq. (\ref{eqn:12}) are different from these in Ref. \cite{Bhattacharyya:2002nc}. Explicitly, these terms cannot exactly resemble what is in the SM, since there are extra contributions from the couplings of the Higgs field with the $A_5$ component in the bulk space as well.

\par The generic structure of the one-loop RGE for the gauge couplings in the five dimensional UED model are given by
\beq
16{\pi ^2}\frac{{d{g_i}}}{{dt}} = \left[{b_i}^{SM} + (S(t) - 1){{\tilde b}_i}\right]{g_i}^3\; , \label{eqn:13}
\eeq
where $b_i^{SM} = \displaystyle \left( \frac{41}{10} , - \frac{19}{6}, -7 \right)$, $\tilde{b}_i = \displaystyle \left( \frac{1}{10} , - \frac{41}{6}, - \frac{21}{2}\right)+ \left( \frac{8}{3} , \frac{8}{3},\frac{8}{3} \right) \eta$, with $\eta$ being the number of generations of matter fields in the bulk. Note that compared with
Ref. \cite{PerezLorenzana:1999qb} we have an additional factor 2 multiplying $\eta$, since in the bulk we need to double count the contributions from both the right- and left-handed KK modes for the closed fermion one-loop diagrams in the wave function renormalization of the gauge fields. Also, due to the compactification having the $Z_2$ symmetry $y \to  - y$, the integral over the extra-dimension is from $0 \to \pi R$, not from $0 \to 2\pi R$, which leads to a factor of two difference with respect to Ref. \cite{Blennow:2011mp}.

\par Therefore, for all our matter fields localized to the 3-brane (that is, $\eta$ = 0), we have
\beq
{\tilde b}_i = \left( \frac{1}{10} , - \frac{41}{6}, - \frac{21}{2}\right) \; , \label{eqn:13.1}
\eeq
where the coupling constant $g_1$ is chosen to follow the conventional SU(5) normalization. These equations (Eqs. (\ref{eqn:88}, \ref{eqn:9}, \ref{eqn:10}, \ref{eqn:11}, \ref{eqn:13})) form a complete set of coupled differential equations for this five dimensional model.


\section{Running of the CKM matrix}\label{sec:3}

\begin{figure}[t]
\begin{center}
\includegraphics[width=0.45\textwidth]{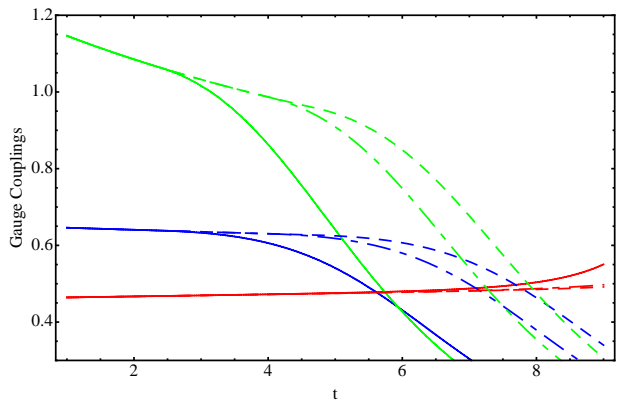}
\caption{\sl (Color online) The evolution of the gauge couplings ($g_1$ red, $g_2$ blue, and $g_3$ green) in the five dimensional brane localized matter fields model, where the solid line is the $R^{-1} = 1~TeV$ case, the dotted-dashed line is the $R^{-1} = 5~TeV$, and the dashed line is the $R^{-1} = 10~TeV$ case.}
\label{fig:1}
\end{center}
\end{figure}

\par As already discussed, extra dimensional models make an interesting $TeV$ scale physics scenario and might be revealed in higher energy collider experiments, as they feature states that have full access to the extended spacetime manifold. For the case of brane localized matter fields, only the boson fields (the gauge fields and the scalar fields) can propagate in the bulk space. However, if the compactification radius $R$ is sufficiently large, due to the power law running of the gauge couplings, it will enable us to bring the unification scale down to an exportable range at the LHC scale. In Fig.\ref{fig:1}, in the regime of the LHC, we plot the evolutions of the gauge couplings in this five dimensional model for different compactification radii of $R^{-1} = 1~TeV$, $5~TeV$, and $10~TeV$ respectively. Here we take the same input  parameters at the EW scale as in Ref. \cite{Cornell:2010sz}. Due to the cumulative effect from each KK level's contributions to the gauge beta function, we observe the fast running of the gauge fields, which quickly lead to the unification scale at $5.7~TeV$, $7.25~TeV$, and $8~TeV$ respectively.

\begin{figure}[t]
\begin{center}
\includegraphics[width=0.45\textwidth]{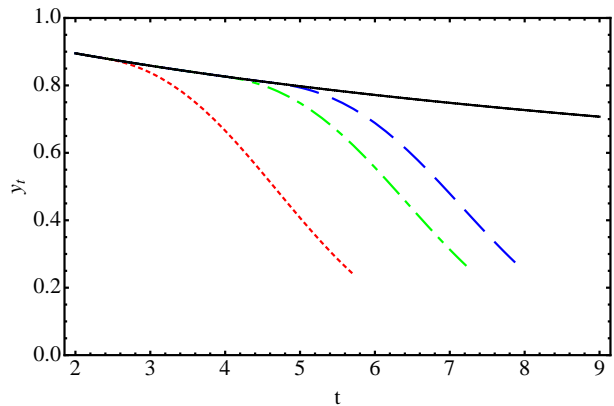}
\caption{\sl (Color online) The evolution of the top Yukawa coupling, where the solid (black) line is the SM, the dotted (red) line is the $R^{-1} = 1~TeV$ case, the dotted-dashed (green) line is the $R^{-1} = 5~TeV$ case, and the dashed (blue) line is the $R^{-1} = 10~TeV$ case.}
\label{fig:2}
\end{center}
\end{figure}

\par On the other hand, due the large hierarchy of the quark masses among different families \cite{Kuo:2005jt}, the top Yukawa coupling is most significant in the beta functions, and it plays a leading role in the evolution. As depicted in Fig.\ref{fig:2}, when the energy of the system is less than the excitations of the first KK modes, the theory follows the evolution of the usual four dimensional SM, and the existence of the KK modes is thus ignored. However, once the first KK threshold is reached, that is when $\mu > R^{-1}$, the contributions from the KK states become more and more significant, and the running deviates from its normal SM orbits and begins to evolve with a faster rate. Consequently, as observed in Fig.\ref{fig:2}, the top Yukawa coupling quickly drops to very small values and terminates at the gauge unification scales. This fast running behavior thus indicates a way to solve the fermion mass hierarchy problem, and the result is also consistent with that in the five dimensional UED model \cite{Cornell:2010sz}.

\par
\begin{figure}[th]
\begin{center}
\epsfig{file=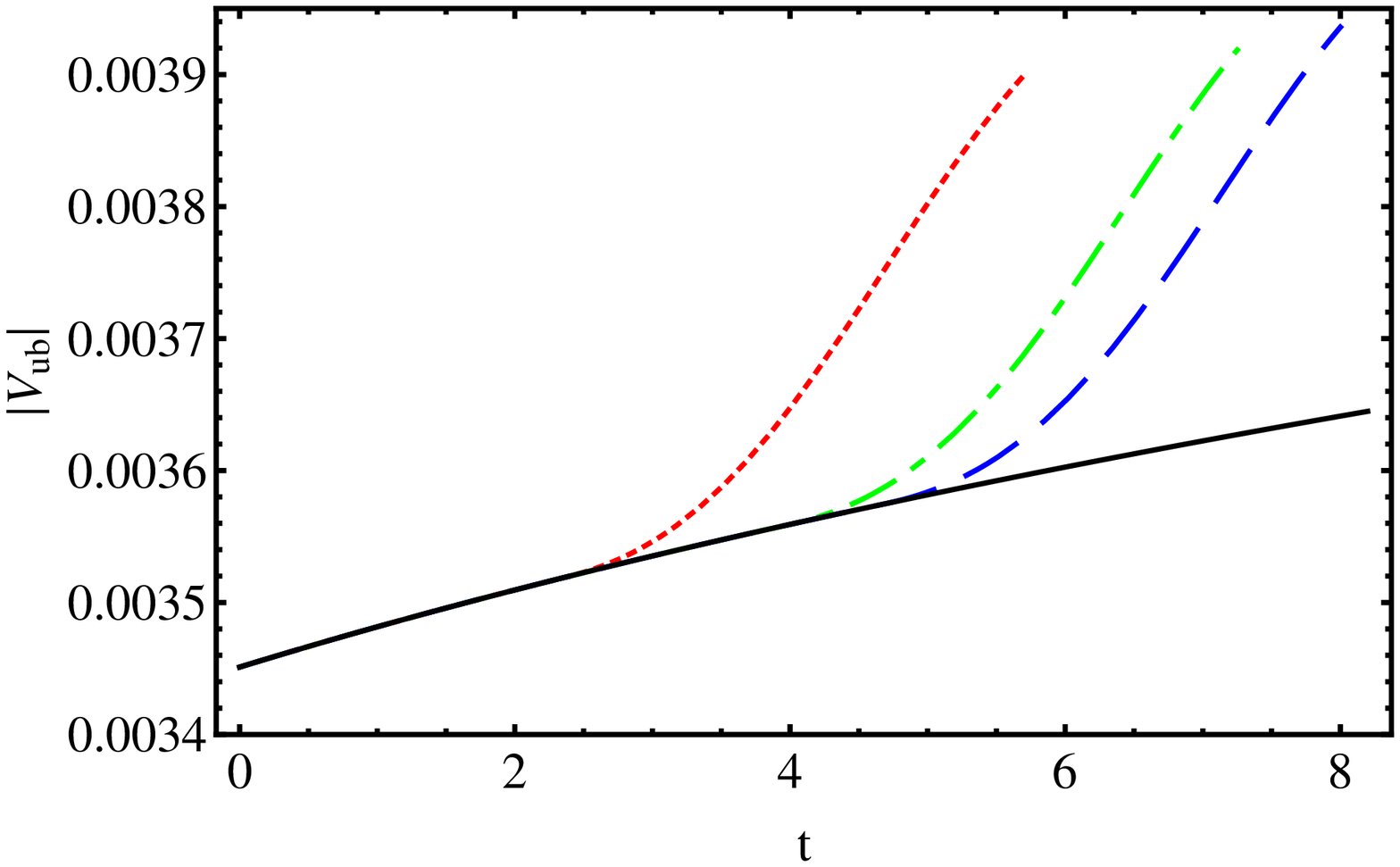,width=.4\textwidth}
\epsfig{file=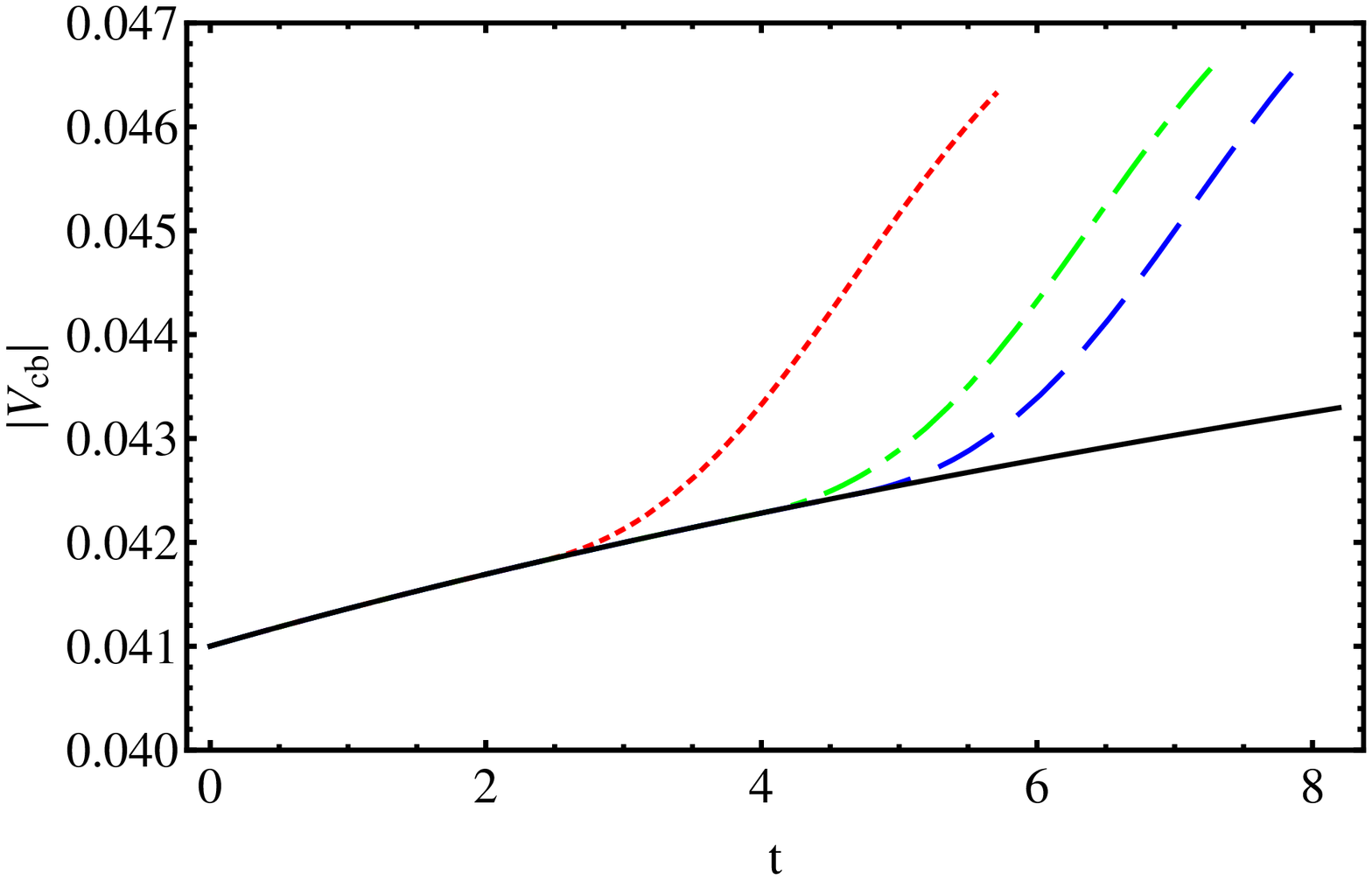,width=.4\textwidth}
\epsfig{file=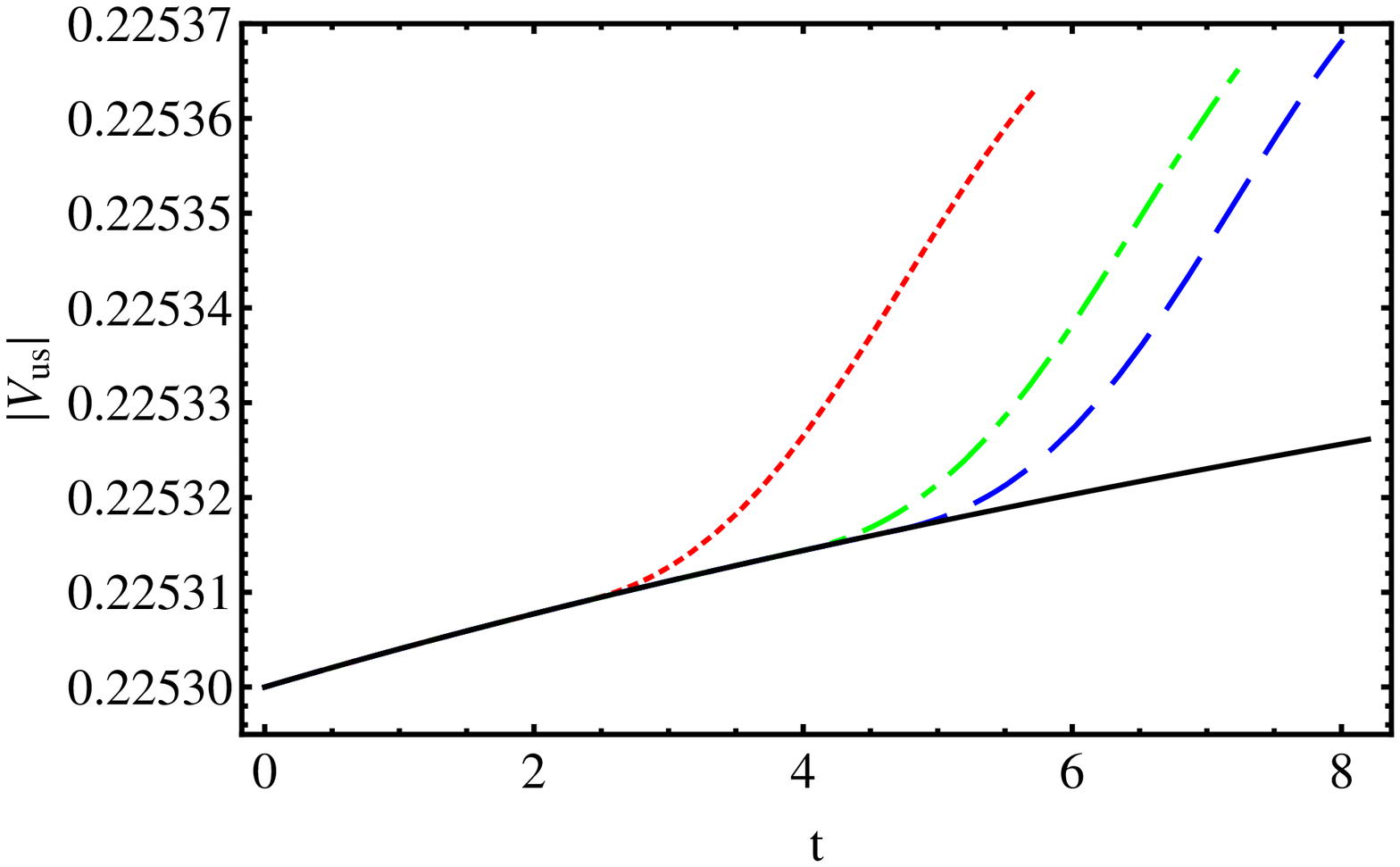,width=.4\textwidth}
\epsfig{file=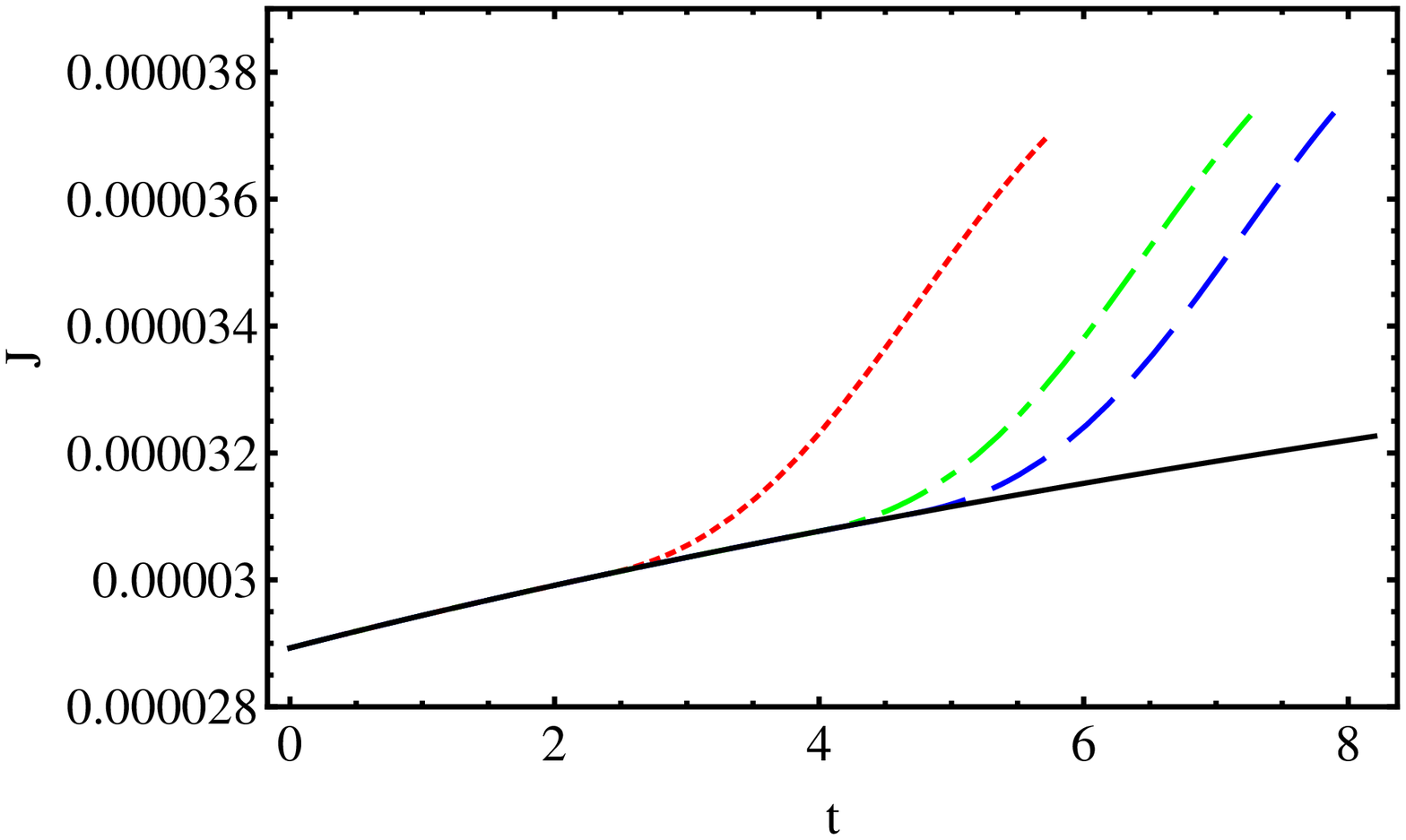,width=.4\textwidth}
\caption{\sl (Color online) The evolution of the CKM matrix element $|V_{ub}|$, top left panel, $|V_{cb}|$ top right panel, $|V_{us}|$ bottom left panel, and $J$, bottom right panel; where the solid line is the SM, the dotted (red) line is the $R^{-1} = 1$ TeV UED case, the dotted-dashed green) line is the 5 TeV UED case and the dashed (blue) line is the 10 TeV UED case.}
\label{fig:3-6}
\end{center}
\end{figure}

 We next turn our attention to the quark flavor mixings. Of the nine elements of the CKM matrix, only four of them are independent, which is consistent with the four independent variables of the standard parameterization of the CKM matrix. For definiteness, we choose the $|V_{ub}|$, $|V_{cb}|$, $|V_{us}|$ and the Jarlskog rephasing invariant parameter $J = \mathrm{Im} V_{ud}V_{cs}V^*_{us} V^*_{cd}$ as the four independent parameters of $V_{CKM}$. In Fig.\ref{fig:3-6} we plot the energy dependence of these four variables from the weak scale all the way up to the unification scale for different values of compactification radii $R$. We observe from these plots the following; in the standard parametrization the CKM matrix elements $V_{ub} \simeq \theta_{13} e^{-i \delta}$, $V_{cb} \simeq \theta_{23}$, can be used to observe the mixing angles $\theta_{13}$ and $\theta_{23}$ and that they increase with the energy scale; the variation rate becoming faster once the KK threshold is passed. For the mixings related to the third family, the UED effects become sizable and the mixing angles $\theta_{13}$ and $\theta_{23}$ change more significantly than the variation of the Cabibbo angle, which appears to be the least sensitive. For the parameter $J$, the characteristic parameter for the CP non-conservation effects, its variation becomes significant. As can be seen from the following section, in light of the current low Higgs mass at the $m_Z$ scale, the evolution of the CKM matrix is valid from the EW scale all the way to the gauge unification scale. In contrast, with the SM and the UED model, the evolution of the CKM matrix becomes unreliable as the new ultraviolet cutoff will arise from the vacuum stability condition when we start from a low value for the Higgs mass.

\section{Improved vacuum stability}\label{sec:4}

\par The value of the Higgs mass itself can provide an important constraint on the scale up to which the SM remains successful as an effective theory. In the SM, the Higgs boson mass is given by $m_H = \sqrt{\lambda}v$, where $\lambda$ is the Higgs self-coupling parameter and $v$ is the vacuum expectation value of the Higgs field (where $v =(\sqrt{2} G_F)^{-1/2} = 246 GeV$ is fixed by the Fermi coupling $G_F$).  From the requirement that the scalar potential energy of the vacuum be bounded from below, the quartic coupling $\lambda$ should be positive at any energy scale. If $m_H$ is too small, $\lambda$ becomes negative at certain energy scales and then induces a false and deep minimum at large field values, destabilizing the EW vacuum. Therefore, above that scale, the validity of the SM is expected to fail and must be embedded in some more general theories that give rise to a wealth of new physics phenomena. In the five dimensional UED model, by using the running of the Higgs self coupling and applying its vacuum stability and singularity conditions, we have derived the correlations between the Higgs mass and the compactification scales \cite{Cornell:2011ge}. There, in terms of the new ATLAS and CMS Higgs bounds around $125GeV$, the scalar potential stability condition will give us a finite ultraviolet cut-off which is much lower than the gauge unification scale.

\par
\begin{figure}[t]
\begin{center}
\includegraphics[width=0.45\textwidth]{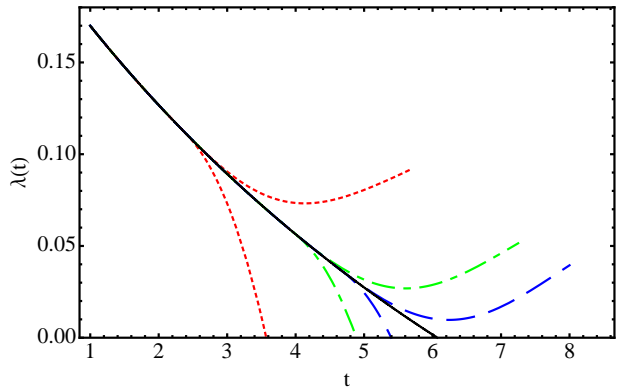}
\caption{\sl (Color online) The evolution of the Higgs self coupling for $m_H = 115.5~GeV$, where the solid (black) line is the SM, the upward [downward] dotted (red) line is the $R^{-1} = 1~TeV$ five dimensional case [UED case], the upward [downward] dotted-dashed (green) line is the $R^{-1} = 5~TeV$ five dimensional case [UED case], and the upward [downward] dashed (blue) line is the $R^{-1} = 10~TeV$ five dimensional case [UED case].}
\label{fig:7}
\end{center}
\end{figure}

\begin{figure}[t]
\begin{center}
\includegraphics[width=0.45\textwidth]{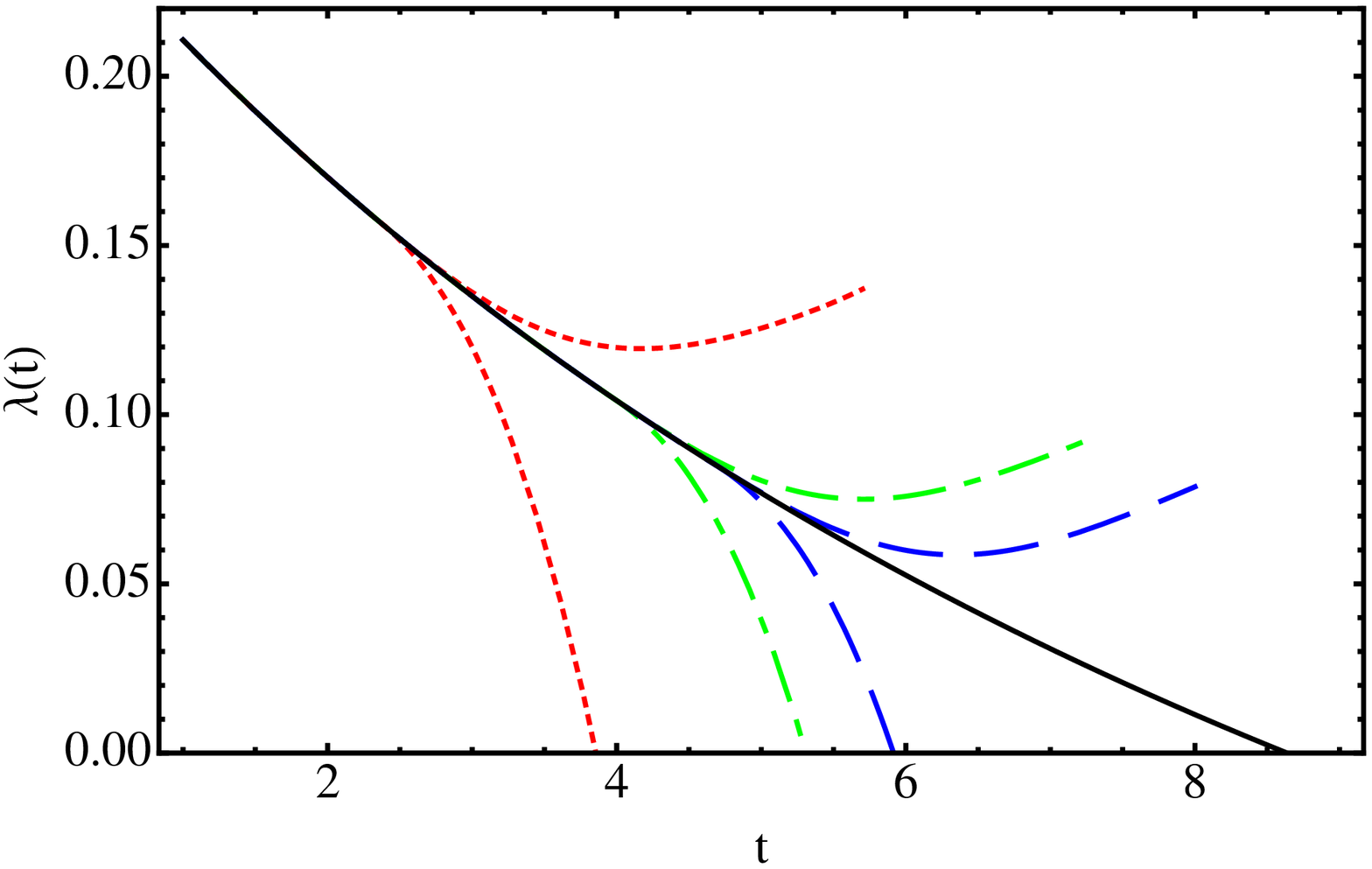}
\caption{\sl (Color online) The evolution of the Higgs self coupling for $m_H = 125 GeV$, where the solid (black) line is the SM, the upward [downward] dotted (red) line is the $R^{-1} = 1 TeV$ five dimensional case [UED case], the upward [downward] dotted-dashed (green) line is the $R^{-1} = 5 TeV$ five dimensional case [UED case], and the upward [downward] dashed (blue) line is the $R^{-1} = 10 TeV$ five dimensional case [UED case].}
\label{fig:8}
\end{center}
\end{figure}

\begin{figure}[t]
\begin{center}
\includegraphics[width=0.45\textwidth]{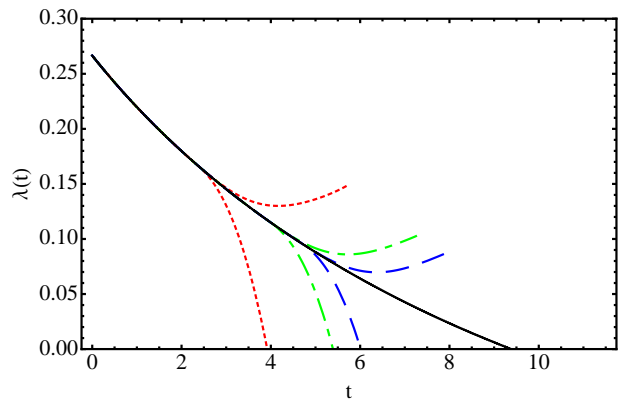}
\caption{\sl (Color online) The evolution of the Higgs self coupling for $m_H = 127 GeV$, where the solid (black) line is the SM, the upward [downward] dotted (red) line is the $R^{-1} = 1 TeV$ five dimensional case [UED case], the upward [downward] dotted-dashed (green) line is the $R^{-1} = 5 TeV$ five dimensional case [UED case], and the upward [downward] dashed (blue) line is the $R^{-1} = 10 TeV$ five dimensional case [UED case].}
\label{fig:9}
\end{center}
\end{figure}

 Following the current context, for the brane localized matter fields scenario, we shall now investigate how the evolution of $\lambda$ depends on the initial values of $\lambda (M_Z)$. We focus on the evolution of the Higgs self-coupling and explore its behavior and constraints on the compactification radii for which the validity of the theory is satisfied. The admissible values of $\lambda (M_Z)$ can be transformed into the allowed values of the Higgs boson mass at $M_Z$. In terms of the current constraints on the Higgs mass we have a range of $115.5~GeV -127~GeV$ from the LHC, for a given initial value of $\lambda (M_Z)$, we follow the differential equation Eq.(\ref{eqn:11}) to pursue its evolution for different compactification radii. For definiteness in Figs.\ref{fig:7}--\ref{fig:9} we illustrate these alternative evolution trajectories for $\lambda (t)$ with different initial values, which are reformulated by using the Higgs mass $m_H (M_Z)$.

\par Specifically, in Fig.\ref{fig:8}, starting from the initial value of $m_H = 125 GeV$, we explore the evolution of $\lambda (t)$. We can observe that, in the whole range from the EW scale up to the gauge unification scale, in this five dimensional model the Higgs self coupling $\lambda (t)$  remains positive and its trajectory goes upward but remains finite when approaching the unification scale, whilst for the SM and the UED model $\lambda (t)$ evolves towards a zero value before reaching the unification scale, which then incurs the vacuum instability and introduces an ultraviolet cutoff for the theory. With a fixed $m_H$, the scaling of $\lambda (t)$ is sensitive to the top quark Yukawa coupling and the strong coupling constant. In fact, as we can see in Eq.(\ref{eqn:12}), the beta function of the Higgs quartic coupling $\lambda$ receives positive contributions from the radiative corrections of the pure scalar and pure gauge fields, and the KK mode loop contributions of these gauge fields to the Higgs self coupling's beta function are also constructive as opposed to the fermion's contributions in the beta function of the SM. Thus, eventually, the quartic terms of the gauge fields in Eq.(\ref{eqn:12}) will overwhelm and overcome the negative contribution from the Yukawa couplings,  and compel the beta function into becoming positive. Qualitatively, the function $\lambda (t)$ will initially fall with energy scale, however, when the first KK threshold is passed, the contributions from the higher KK modes of the gauge fields become more and more significant, finally overcoming the negative contribution from the Yukawa couplings and then raising the Higgs self coupling, which then helps to stabilize the EW vacuum up to a high scale.

\par In contrast, in the UED model both the fermion fields and the boson fields can propagate in the bulk. For a small $m_H$, like the current LHC Higgs bound, we find the case of the vacuum stability being very sensitive to the contributions of Yukawa couplings. In Eq.(\ref{eqn:A7}) the Yukawa coupling terms $(Y^\dag Y)^2$ dominate and have a negative contribution to the beta function. This causes the scalar self coupling to decrease and reach zero at a certain energy scale. As a result, it can destabilize the vacuum at an energy scale far below the unification scale. Compared with the SM case, in the UED model, once the first KK threshold is passed, the KK fermion contributions are sizable and make the trajectory evolve downward and quickly go to zero. In order to rescue the stability of the vacuum, it is necessary to introduce new physics at such a scale which would have a non-negligible impact on the radiative corrections to the scalar potential and raise it.

\par In Figs.\ref{fig:7} and \ref{fig:9}, we analyze the vacuum stability for the lower and upper bounds of the Higgs mass from the LHC. As observed before, for an even smaller Higgs mass, even though the Higgs self coupling is positive at a low energy scale, for the case of the SM and UED model, it evolves towards zero at a faster rate at higher energies, signifying the necessity of introducing an ultraviolet cutoff. In Fig.\ref{fig:9}, a relatively large Higgs mass has a milder evolution behavior, but is still outside the range of the Higgs mass which would lead to a finite value at a high energy scale \cite{Cornell:2011ge}.

\section{Summary}\label{sec:4}

\par In summary, in this work we have investigated the evolution of the quark flavor mixings and the scalar self coupling for a five dimensional brane localized matter fields model, and analyzed its effects and patterns for different compactification radii $R$. While the evolution of the Cabibbo angle is tiny, the mixing behaviors involving the third generation are sizable. The contribution of KK modes to the scaling of $J$ is also substantial. We have also investigated the impact of the recent ATLAS and CMS Higgs bounds on the renormalization group running of our model. For interesting values of a Higgs mass around $125~GeV$, in contrast with what has been observed in the SM and five dimensional UED model, we find that the Higgs self coupling evolution has a finite value, thus there is no vacuum stability concerns, all the way up to the unification scale.


\appendix

\section{}\label{sec:A}

\par In comparison with the five dimensional brane localized matter fields scenario, for completeness, we list the evolution equations for the five dimensional UED case \cite{Cornell:2011ge,Cornell:2010sz}. The one-loop RGE for the Yukawa couplings have the following form
\bea
16{\pi ^2}\frac{{d{Y_U}}}{{dt}} &=& \beta _U^{SM} + \beta _U^{UED}\nonumber \; , \\
16{\pi ^2}\frac{{d{Y_D}}}{{dt}} &=& \beta _D^{SM} + \beta _D^{UED}\nonumber \; , \\
16{\pi ^2}\frac{{d{Y_E}}}{{dt}} &=& \beta _E^{SM} + \beta _E^{UED} \; , \label{eqn:A1}
\eea
in which $Y_E = diag(y_e, y_\mu, y_\tau)$, these Yukawa coupling beta functions are
\bea
\beta _U^{SM} &=& {Y_U}\left\{  - \left(8g_3^2 + \frac{9}{4}g_2^2 + \frac{{17}}{{20}}g_1^2\right) + \frac{3}{2}(Y_U^\dag {Y_U} - Y_D^\dag {Y_D}) + Tr\left[3Y_U^\dag {Y_U} + 3Y_D^\dag {Y_D} + Y_E^\dag {Y_E}\right]\right\} \nonumber \; , \\
\beta _U^{UED} &=& {Y_U}(S(t) - 1)\left\{ - \left(\frac{{28}}{3}g_3^2 + \frac{{15}}{8}g_2^2 + \frac{{101}}{{120}}g_1^2\right) + \frac{3}{2}(Y_U^\dag {Y_U} - Y_D^\dag {Y_D}) + 2(S(t) - 1)Tr\left[3Y_U^\dag {Y_U} + 3Y_D^\dag {Y_D} + Y_E^\dag {Y_E}\right]\right\} \nonumber \; , \\
\beta _D^{SM} &=& {Y_D}\left\{  - \left(8g_3^2 + \frac{9}{4}g_2^2 + \frac{1}{4}g_1^2\right) + \frac{3}{2}(Y_D^\dag {Y_D} - Y_U^\dag {Y_U}) + Tr\left[3Y_U^\dag {Y_U} + 3Y_D^\dag {Y_D} + Y_E^\dag {Y_E}\right]\right\} \nonumber \; , \\
\beta _D^{UED} &=& {Y_D}(S(t) - 1)\left\{ - \left(\frac{{28}}{3}g_3^2 + \frac{{15}}{8}g_2^2 + \frac{{17}}{{120}}g_1^2\right) + \frac{3}{2}(Y_D^\dag {Y_D} - Y_U^\dag {Y_U}) +  + 2(S(t) - 1)Tr\left[3Y_U^\dag {Y_U} + 3Y_D^\dag {Y_D} + Y_E^\dag {Y_E}\right]\right\} \nonumber \; , \\
\beta _E^{SM} &=& {Y_E}\left\{ Tr\left[3Y_U^\dag {Y_U} + 3Y_D^\dag {Y_D} + Y_E^\dag {Y_E}\right] - \left(\frac{9}{4}g_2^2 + \frac{9}{4}g_1^2\right) + \frac{3}{2}Y_E^\dag {Y_E}\right\} \nonumber \; , \\
\beta _E^{UED} &=& {Y_E} (S(t) - 1)\left\{ - \left(\frac{{15}}{8}g_2^2 + \frac{{99}}{{40}}g_1^2\right) + \frac{3}{2}Y_E^\dag {Y_E}] + 2(S(t) - 1)Tr\left[3Y_U^\dag {Y_U} + 3Y_D^\dag {Y_D} + Y_E^\dag {Y_E}\right]\right\} \; . \label{eqn:A2}
\eea
Explicitly, after diagonalizing the square of the quark Yukawa coupling matrices, we have the RGEs for the $f_i^2$ and $h_j^2$ (these being the eigenvalues of $Y_U^\dag Y_U$ and $Y_D^\dag Y_D$ respectively)
\bea
16{\pi ^2}\frac{{d{f_i}^2}}{{dt}} &=& {f_i}^2\left[2(2S(t) - 1)T - 2{G_U} + 3S(t){f_i}^2 - 3S(t)\sum\limits_j {{h_j}^2} {\left| {{V_{ij}}} \right|^2}\right]\nonumber \; , \\
16{\pi ^2}\frac{{d{h_j}^2}}{{dt}} &=& {h_j}^2\left[2(2S(t) - 1)T - 2{G_D} + 3S(t){h_j}^2 - 3S(t)\sum\limits_i {{f_i}^2} {\left| {{V_{ij}}} \right|^2}\right]\; , \label{eqn:A3}
\eea
and for the lepton sector we have
\beq
16{\pi ^2}\frac{{d{y_a}^2}}{{dt}} = {y_a}^2\left[2(2S(t) - 1)T - 2{G_E} + 3S(t){y_a}^2\right]\; , \label{eqn:A4}
\eeq
where $T = Tr\left[ 3 Y_U^\dag Y_U + 3 Y_D^\dag Y_D + Y_E^\dag Y_E \right]$, $G_U = \displaystyle 8 g_3^2 + \frac{9}{4} g_2^2 + \frac{17}{20} g_1^2 + \left( S(t) - 1 \right) \left( \frac{28}{3} g_3^2 + \frac{15}{8} g_2^2 + \frac{101}{120} g_1^2 \right)$,\\ $G_D = \displaystyle 8 g_3^2 + \frac{9}{4} g_2^2 + \frac{1}{4} g_1^2 +  \left( S(t) - 1 \right) \left( \frac{28}{3} g_3^2 + \frac{15}{8} g_2^2 + \frac{17}{120} g_1^2 \right)$, and $G_E = \displaystyle \left(  \frac{9}{4} g_2^2 + \frac{9}{4} g_1^2 \right) + \left( S(t) - 1 \right) \left( \frac{15}{8} g_2^2 + \frac{99}{40} g_1^2 \right)$.

\par The running of the quark flavor mixing matrix in the charged current is governed by
\bea
16{\pi ^2}\frac{{d{{\left| {{V_{ij}}} \right|}^2}}}{{dt}} &=& S(t)\left\{ {3{{\left| {{V_{ij}}} \right|}^2}\left({f_i}^2 + {h_j}^2 - \sum\limits_k {{f_k}^2} {{\left| {{V_{kj}}} \right|}^2} - \sum\limits_k {{h_k}^2} {{\left| {{V_{ik}}} \right|}^2}\right)} \right. \nonumber \\
&& \hspace{1cm} - 3{f_i}^2\sum\limits_{k \ne i} {\frac{1}{{{f_i}^2 - {f_k}^2}}} \left(2{h_j}^2{\left| {{V_{kj}}} \right|^2}{\left| {{V_{ij}}} \right|^2} + \sum\limits_{l \ne j} {{h_l}^2{V_{iklj}}} \right) \nonumber \\
&& \hspace{1cm} - 3{h_j}^2\sum\limits_{l \ne j} {\frac{1}{{{h_j}^2 - {h_l}^2}}} \left(2{f_i}^2{\left| {{V_{il}}} \right|^2}{\left| {{V_{ij}}} \right|^2} + \sum\limits_{k \ne i} {{f_k}^2{V_{iklj}}} \right)\; , \label{eqn:A5}
\eea
and the beta function of the Higgs self coupling is given by
\beq
16{\pi ^2}\frac{{d\lambda }}{{dt}} = \beta _\lambda ^{SM} + \beta _\lambda ^{UED}\; , \label{eqn:A6}
\eeq
where
\bea
\beta _\lambda ^{SM} &=& 12{\lambda ^2} - \left( {\frac{9}{5}g_1^2 + 9g_2^2} \right)\lambda  + \frac{9}{4}\left( {\frac{3}{{25}}g_1^4 + \frac{2}{5}g_1^2g_2^2 + g_2^4} \right) + 4\lambda Tr[3Y_U^\dag {Y_U} + 3Y_D^\dag {Y_D} + Y_E^\dag {Y_E}] \nonumber \\
&& \hspace{1cm}- 4Tr[3{(Y_U^\dag {Y_U})^2} + 3{(Y_D^\dag {Y_D})^2} + {(Y_E^\dag {Y_E})^2}]\nonumber \; , \\
\beta _\lambda ^{UED} &=& (S(t) - 1)\left\{ {12{\lambda ^2} - 3\left( {\frac{3}{5}g_1^2 + 3g_2^2} \right)\lambda  + \left( {\frac{9}{{25}}g_1^4 + \frac{6}{5}g_1^2g_2^2 + 3g_2^4} \right)} \right\} \nonumber \\
&& + 2(S(t) - 1)\left\{ {4\lambda Tr[3Y_U^\dag {Y_U} + 3Y_D^\dag {Y_D} + Y_E^\dag {Y_E}] - 4Tr[3{{(Y_U^\dag {Y_U})}^2} + 3{{(Y_D^\dag {Y_D})}^2} + {{(Y_E^\dag {Y_E})}^2}]} \right\}\; . \label{eqn:A7}
\eea

\par Furthermore, recall that the general beta functions of the one-loop RGE for the gauge couplings have the form
\beq
16{\pi ^2}\frac{{d{g_i}}}{{dt}} = \left[{b_i}^{SM} + (S(t) - 1){{\tilde b}_i}\right]{g_i}^3\; , \label{eqn:A8}
\eeq
where $b_i^{SM} = \displaystyle \left( \frac{41}{10} , - \frac{19}{6}, -7 \right)$, $\tilde{b}_i = \displaystyle \left( \frac{81}{10} , \frac{7}{6}, - \frac{5}{2}\right)$. These equations (Eqs.(\ref{eqn:A1}--\ref{eqn:A8})) form a complete set of coupled differential equations for the three generations of fermions for the UED model.


\end{document}